\documentclass[aps,pra,reprint,groupedaddress,showpacs,10pt]{revtex4-1}

\usepackage{mathrsfs}
\usepackage{amsfonts}
\usepackage{amsmath} %Math commands 
\usepackage{amssymb} %Extra symbols 
\usepackage{amsthm}
\usepackage{graphicx} %Include postscript graphics 
\usepackage{url}
\usepackage{color}
\usepackage{hyperref}
\definecolor{refcolor}{RGB}{0,0,190}
\hypersetup{
    colorlinks,
    citecolor=refcolor,
    filecolor=refcolor,
    linkcolor=refcolor,
    urlcolor=refcolor
}

\begin{document}

%\preprint{}

%\title{Lost in the Hilbert space}
\title{Can the clicks of the detectors provide a complete description of Nature?}


\author{Cristinel Stoica}
\email[]{cristi.stoica@theory.nipne.ro}
%\thanks{}
\affiliation{Department of Theoretical Physics, \\Horia Hulubei National Institute for Physics and Nuclear Engineering, 077125, Bucharest, Romania.}

%\email[]{}


\date{\today}

\begin{abstract}
No matter how counterintuitive they are, quantum phenomena are all simple consequences of the laws of Quantum Mechanics. It is not needed to extend the theory with hidden mechanisms or additional principles to explain what Quantum Mechanics already predicts.

This indubitable fact is often taken as supporting the view that all we can know about the universe comes from the outcomes of the quantum observations. According to this view, we can even  learn the physical laws, in particular the properties of the space, particles, fields, and interactions, solely from the outcomes of the quantum observations.

In this article it is shown that the unitary symmetry of the laws of Quantum Mechanics imposes severe restrictions in learning the physical laws of the universe, if we know only the observables and their outcomes.
\end{abstract}


%\keywords{}

\maketitle

\tableofcontents


\theoremstyle{definition}
\newtheorem{theorem}{Theorem}%[section]
\newtheorem{proposition}[theorem]{Proposition}
\newtheorem{lemma}{Lemma}
\newtheorem{remark}{Remark}
\newtheorem{corollary}{Corollary}
\newtheorem{definition}{Definition}
\newtheorem{principle}{Principle}
\newtheorem{hypothesis}{Hypothesis}
\newtheorem{property}{Property}

%--------------------------------------------------------
% Defines
   
\def\dirlim #1{\lim_{\genfrac{}{}{0pt}{}{\longrightarrow}{#1}}}

\def\({\left(}
\def\){\right)}

\def\ntto{\to}

\newcommand{\tn}{\textnormal}
\newcommand{\ds}{\displaystyle}
\newcommand{\dsfrac}[2]{\displaystyle{\frac{#1}{#2}}}

\newcommand{\mc}[1]{\mathcal{#1}}
\newcommand{\ms}[1]{\mathscr{#1}}
\newcommand{\statespace}{\mathcal{S}}
\newcommand{\hilbert}{\mc{H}}
\newcommand{\vectorspace}{\mathcal{V}}

\newcommand{\qmU}{$\mathscr{U}$}
\newcommand{\qmR}{$\mathscr{R}$}
\newcommand{\qmUR}{$\mathscr{UR}$}
\newcommand{\qmDR}{$\mathscr{DR}$}

\newcommand{\R}{\mathbb{R}}
\newcommand{\C}{\mathbb{C}}
\newcommand{\Z}{\mathbb{Z}}
\newcommand{\K}{\mathbb{K}}
\newcommand{\N}{\mathbb{N}}
\newcommand{\Prj}{\mathcal{P}}
\newcommand{\abs}[1]{\left|#1\right|}

\newcommand{\de}{\tn{d}}
\newcommand{\tr}{\tn{tr}}

\newcommand{\ie}{\textit{i.e.} }
\newcommand{\vs}{\textit{vs.} }
\newcommand{\eg}{\textit{e.g.} }
\newcommand{\etc}{\textit{etc}}
\newcommand{\etal}{\textit{et al.}}

\newcommand{\pde}{PDE}
\newcommand{\U}{\tn{U}}
\newcommand{\SU}{\tn{SU}}

\newcommand{\schrod}{Schr\"odinger}
\newcommand{\vonneum}{Liouville - von Neumann}
\newcommand{\ks}{Kochen-Specker}
\newcommand{\leggarg}{Leggett-Garg inequality}
\newcommand{\bra}[1]{\langle#1|}
\newcommand{\ket}[1]{|#1\rangle}
\newcommand{\braket}[2]{\langle#1|#2\rangle}
\newcommand{\expectation}[1]{\langle#1\rangle}
\newcommand{\Herm}{\tn{Herm}}

\newcommand{\Hint}{H_{\tn{int}}}

\newcommand{\quot}[1]{``#1''}

\def\sref #1{\S\ref{#1}}

\newcommand{\Law}[1]{\textbf{#1}}
\newcommand{\RO}{\Law{RO}}
\newcommand{\RP}{\Law{RP}}
\newcommand{\RC}{\Law{RC}}
\newcommand{\TP}{\Law{TP}}

\newcommand{\img}[3]{\begin{figure}[ht]\includegraphics[width=#2\textwidth]{#1.eps}
\caption{\small{\label{#1}#3}}\end{figure}}

\newcommand{\todo}[1]{\textbf{TODO: #1$\P$}\PackageWarning{TODO:}{#1!}}
\newcommand{\tocheck}{\PackageWarning{Check this!!!}{Check this!!!}}

\newcommand{\image}[3]{
%\begin{center}
\begin{figure}[!ht]
\includegraphics[width=#2\textwidth]{#1}
\caption{\small{\label{#1}#3}}
\end{figure}
%\end{center}
}

%---------------------------------------------------------------------------------------------------%
\section{Introduction}

Quantum Mechanics (QM) has a simple formalism \cite{Dir58,vonNeumann1955foundations}, in which the state is a vector in the Hilbert space, the evolution is governed by a unitary operator, the observables are Hermitian operators, and the outcomes are eigenvalues of the observables. The probabilities of the outcomes are given by the Born rule. This formalism predicts phenomena which have no counterpart in classical physics: complementarity, uncertainty, entanglement, correlations in space and time, contextuality, apparent violations of relativity and causality, \etc.

No matter how strange may appear quantum correlations between systems separated in space or time, they follow directly from the postulates of QM. Faster than light signaling is neither required nor allowed. The notion of space, together with those of position and locality, play no role in the explanation of quantum correlations \cite{Sto14QMa}.

According to Bohr, quantum phenomena appear mysterious to us because we try to assign an objective reality between observations. As J.A. Wheeler put it, ``no phenomenon is a phenomenon, until it is an observed phenomenon'' \cite{Whe78}, and according to Asher Peres, ``unperformed experiments have no results'' \cite{Peres1978UnperformedExperimentsHaveNoResult}.
Quantum phenomena are predicted, therefore explained by the projection postulate and the Born rule \cite{Sto14QMa}, but to see this, we have to unlearn our classical intuitions about reality and causality, and embrace

\textbf{The unborn rule}: {\em reality is unborn until is observed}.

Given that quantumness follows from and is explained by the principles of QM, no additional explanation or interpretation is needed.
It is true that QM appears very different from classical physics, but this doesn't mean that we should try to make it more similar to the other theories.

The understanding of this self-sufficiency of QM led notorious physicists to the idea that {\em everything we can learn from the world are outcomes of quantum observations}. They defend the position that the collection of observables, together with the outcomes of the measurements, are all we can get, and this is enough to understand how the universe works. This position has its roots in Bohr's writings, and was developed to its apogee by Wheeler with his ``it from bit'' program \cite{Whe78,wheeler1983recognizing,wheeler1988world,wheeler1989information,Whe98geons}, by Mermin with his (in)famous ``shut up and calculate'' \cite{Mermin1989ShutUpAndCalculate}, and more recently by Fuchs and Peres, who wrote that ``quantum theory needs no {`interpretation'}'' \cite{FuchsPeres2000QMNoInterpretation}.

John Wheeler even proposed a program according to which there is no space, no time, no continuum, not even law, all being just secondary constructs, while the fundamental entities are the ``bits'', the outcomes of the measurements \cite{wheeler1989information}.

According to this view, all we can know is the collection of the observables, and the outcomes of their measurements. And this is all we need to reconstruct the world, and anything else is ``metaphysics''.

In this article, I provide an analysis of the ``a universe from the clicks of the detectors'' viewpoint. I show that the invariance under unitary transformations of the Hilbert space makes quantum theories to be characterized only by the Hilbert space. But any Hilbert spaces of finite or countable number of dimensions is isomorphic to any other Hilbert space of the same number of dimensions, and a Hilbert spaces having uncountable basis can't be distinguished by a finite sequence of measurements from one with a countable or finite basis. The Hilbert space contains no information about space and its dimensions, about particles, fields, interactions \etc. Therefore, there is nothing we can learn about the universe if we rely only on quantum observations.

On the other hand, if we supplement the Hilbert space with additional structures, like tensor product decomposition and a position basis, then not only the space, but also the particles, fields, and interactions emerge.
This suggests that there is something more than the outcomes of the quantum observations, and the fact that quantum fields have a Hilbert space structure may be a consequence, rather than a fundamental principle.

%---------------------------------------------------------------------------------------------------%
\section{\label{s:universe_from_clicks}A universe from the clicks of the detectors}

%---------------------------------------------------------------------------------------------------%
\subsection{\label{s:QMprinciples}The axioms of Quantum Mechanics}

We will rely on the well known postulates of Quantum Mechanics \cite{Dir58,vonNeumann1955foundations}. For simplicity, let's remember a version which I reproduce from Fuchs and Stacey \cite{Fuchs2014NegativeRemarks}:
\begin{enumerate}
	\item Associated with each system is a complex vector space $\hilbert$.
	\item Measurements correspond to orthonormal bases $\ket{e_i}$ on $\hilbert$.
	\item States correspond to density operators $\rho$ on $\hilbert$.
	\item Systems combine by tensor producting their vector spaces, $\hilbert_{AB}=\hilbert_A\otimes\hilbert_B$.
	\item When no measurements are performed, states evolve by unitary maps $\ms U$.
\end{enumerate}

There are some counterintuitive quantum phenomena, manifest in Bohr complementarity, Heisenberg uncertainty \cite{Heisenberg1927Uncertainty,Kennard1927Uncertainty,Weyl1928GruppentheorieQuantenmechanik}, EPR-Bell experiment \cite{EPR35,Bohm51,Bel64}, {\ks} theorem \cite{KochenSpecker1967HiddenVariables}, {\leggarg} theorem \cite{LeggettGarg1985QMvsMacrorealism}. There is no doubt that they follow from the very postulates of QM \cite{Sto14QMa}.

However, can we reconstruct the physical laws of the entire universe from the quantum observations?

%---------------------------------------------------------------------------------------------------%
\subsection{\label{s:QuantumHistory}Histories of quantum observations}

Any observation of a subsystem is in fact an observation of the entire system, although it only gives information about the state of the subsystem.
Consider for instance a quantum system $\mc S$, which may be the entire universe, and its associated Hilbert space $\hilbert$. Let $\mc S_A$ be a subsystem of $\mc S$, having associated the Hilbert space $\hilbert_A$. Then, if $\hilbert_B$ is the Hilbert space of the rest of the system, $\hilbert=\hilbert_A\otimes\hilbert_B$. An observable $\mc{\hat O}$ of $\mc S_A$ can be considered to be an observable of the entire system  $\mc S$, because it can be extended to the entire Hilbert space $\hilbert$ by taking the tensor product $\mc{\hat O}\otimes I_B$ of $\mc{\hat O}$  and the identity operator on $\hilbert_B$. Therefore, we can consider without loss of generality that all observables of subsystems are also observables of the entire system.

The Hilbert space is invariant under unitary transformations. A unitary transformation maps observables to other observables. Moreover, all the axioms described in \sref{s:QMprinciples} are invariant to unitary transformations. Let's call this principle that axioms of Quantum Mechanics are invariant under unitary transformations, for future reference, the {\em principle of unitary invariance}.

This invariance lies at the root of the equivalence between the Heisenberg and the {\schrod} pictures. We will work in the Heisenberg picture, therefore considering that the state vector is constant between observations. The observables under consideration are obtained from the observables in the {\schrod} picture by applying a unitary transformation given by the unitary evolution operator.

We arrive at the conclusion that the observable history of a quantum system, which can be the entire universe, is described by a sequence of pairs
\begin{equation}
\label{eq:QuantumHistory}
	(\mc{\hat O}_i,\lambda_i)_{i\in A},
\end{equation}
where each $\mc{\hat O}_i$ is an observable, and $\lambda_i$ the outcome obtained by measuring it. The set $A$ indexing the measurements can be for example the set of integers, $A=\Z$.

In the following I will address the following question:

{\em Is it possible to learn something about the laws governing the universe, solely from a history of quantum measurements \eqref{eq:QuantumHistory}?}

I will argue that it is not possible to learn very much about our world merely by the outcomes of the measurements. In fact, the only information we can obtain from a history of quantum observations is that the dimension of the Hilbert space is a big number, possibly infinite.

%---------------------------------------------------------------------------------------------------%
\section{What can we really learn from the quantum observations?}

%---------------------------------------------------------------------------------------------------%
\subsection{The forgetful quantization}

The standard way to obtain a quantum theory is to start with a classical one and quantize it.

The classical theories from which the quantum theory is obtained can be very diverse. For example, space can have any number of dimensions, it can be continuous or discrete, it may even be no space as we conceive it at all, there are no restrictions on the kinds of particles, fields and the evolution equations they obey \etc. No matter how the classical theory we quantize is, the result is always a quantum theory obeying the principles of QM. Can we recover the classical theory, by knowing its quantized version? How can we even know that a given quantum theory is obtained by quantizing a classical one?

No matter how we obtain the quantum theory, at the end we remain only with a Hilbert space and a Hamiltonian, which are subject to unitary invariance. As I will explain below, this means that from the principles of QM we can't say much about our world. They don't contain information about the number of dimensions and even about the existence of space, neither about the kinds of particles, fields and the evolution equations they follow.

%---------------------------------------------------------------------------------------------------%
\subsection{The Hilbert space is too symmetric}

One may think of course that after quantization, there will still be information about the number of space dimensions, as well as the spin and the internal degrees of freedom, encoded in the parameters which index the state vectors in the Hilbert space. Also, the dynamics will be contained in the Hamiltonian generating the unitary evolution, and the interactions will be visible once we express the Hilbert space as a tensor product of smaller spaces, corresponding to elementary particles.

This would be true, if this information could be extracted merely from the history of measurements from equation \eqref{eq:QuantumHistory}. In the {\schrod} picture, the evolution of the state of a system can be described as the unitary rotation of a vector in the Hilbert space, interrupted from time to time by jumps from one state to another, according to the projection postulate and the Born rule. The Hilbert space doesn't have a preferred basis, in the sense that the principles of QM are independent on the basis, and we can even choose it so that it rotates unitarily. For instance, we can go from the {\schrod} picture to the Heisenberg picture, by applying to the Hilbert space the unitary transformation inverse to the unitary evolution operator, so that the state vector remains constant in time. According to the Heisenberg picture, the state vector is constant between observations, and the observables are transformed by the unitary operator, but they still are observables. The Born rule remains valid, and the evolution of the system translates as jumps of the state vector from one constant value to another one.

Any sequence of observables is allowed in principle, and any outcomes of them. Therefore we see that, at least in the Heisenberg picture, there is no information about the universe, other than the Hilbert space structure.

%---------------------------------------------------------------------------------------------------%
\subsection{Lost in the Hilbert space}

We have seen that {\em any quantum theory in the Heisenberg picture is completely determined by the Hilbert space}. But if the Hilbert space is finite dimensional, it is isomorphic to any other Hilbert space of the same dimension. Similarly, if it is infinite-dimensional and admits a countable basis, \ie if it is {\em separable}, it is isomorphic to any other infinite-dimensional separable Hilbert space. And if it is not separable, \ie if its basis is not countable, it would be impossible to distinguish it by a countable number of observations from a separable Hilbert space. A quantum theory looks the same as any other quantum theory having a Hilbert space of the same dimension.

It follows that {\em any quantum theory is completely determined by the dimension of the Hilbert space}, in the finite or infinite but countable case. The uncountable case is indistinguishable by experiments from a countable one, and practically even from a finite one with a very large number of dimensions. Moreover, it can always turn out that we observed so far just a subsystem, and the Hilbert space is in fact larger.

{\em How can we then learn something about the universe only from the quantum theory describing it, when this is completely determined by a number which we can't even know after a finite number of observations?}

%---------------------------------------------------------------------------------------------------%
\section{What are the clicks of the detectors not telling us?}

We arrived at the conclusion that {\em quantum observations are not enough}. If we rely solely on the structure of the Hilbert space, we can't learn anything about the universe only from the sequence of observations and their outcomes. Such a quantum universe would be characterized only by the dimension of the Hilbert space, which doesn't tell anything about space, particles, fields, spin and other features of our rich universe. Not to mention the emergence of the classical world, the coexistence with Relativity (which suggests a local and causal universe, at least as an effective limit) and the quantization of gravity.
This suggests that the Hilbert space structure may not be fundamental, or at least the description of the physical world it provides is incomplete.

The only way to establish a connection between the quantum outcomes and the physical world is by {\em supplementing the quantum theory with something else}.

For example, we can supplement the Hilbert space with a preferred tensor product decomposition, \ie $\hilbert = \otimes_\alpha\hilbert_\alpha$, where the Hilbert spaces $\hilbert_\alpha$ correspond to individual particles. We can do better than this, and use instead Fock spaces, for many particles of the same type.

Then, we can supplement the Hilbert space of a single particle with a preferred basis representing the position, or rather a class of position bases related by rotations and translations. 
It is even possible that the Hamiltonian helps identifying a basis of positions, because the interactions contained in it are local. This may be true, but in order to recover local interactions from the Hamiltonian, we have to know a prefered tensor product decomposition of the Hilbert space, because the interactions are between particles.
From the position bases we can obtain by Fourier transform the momentum bases. When expressing a state vector in the Hilbert space of a single particle in a position basis, it becomes a function depending on the position. If in addition to the position there are other degrees of freedom, they will give us the spin and the internal degrees of freedom corresponding to the fundamental forces.

Now, since we have singled out in the theory both the one-particle Hilbert spaces and the positions and momenta, when expressing the Hamiltonian with respect to the tensor product decompositions, the positions, the momenta, and the gauge degrees of freedom, the interactions between particles become explicit.

It follows that {\em the only way to extract from a quantum theory relevant information about the physical world is to supplement it with additional structures}.

Interestingly, Bohr insisted on maintaining the dichotomy between the classical apparatus and the quantum systems to be observed. If we keep the world divided in a classical part, which performs the measurement, and a quantum part, which is measured, then the classical part provides the needed information to supplement the history of quantum observations from equation \eqref{eq:QuantumHistory}.
The classical part from Bohr's interpretation provides the physical meaning of the Hilbert space, the observables, and the outcomes.

One can argue that the {\em decoherence program} \cite{zurek1981pointer,Zur03a} solves the problem of the emergence of the classical world from the quantum one. Let's assume that it does, although this hope is severly limited by the Leggett-Garg theorem \cite{LeggettGarg1985QMvsMacrorealism,Leggett2008Realism}, which rules out macroscopic realism.
However, in order to prove the emergence of the classical from the quantum, the decoherence program relies heavily on an environment which induces the selection of a preferred basis, by having itself properties very close to the ones which are supposed to emerge. The toy examples of environment-induced decoherence assume that the environment is already a separable state \cite{Zur03a}, hence it is very classical, so it is no wonder that it induces the decoherence of a quantum system with much fewer degrees of freedom. Therefore, decoherence explains why systems that behave more classically make other systems decohere, like in the measurement process, but how the classicality of these systems emerged in the first place remains a mystery.

In conclusion, if we try to remove completely the classical part, and rely solely on the clicks in the detectors, we get lost in the Hilbert space.

%\begin{acknowledgments}
%The author cordially thanks 
%xxx, 
%for very helpful comments and suggestions.
%\end{acknowledgments}

%\bibliographystyle{unsrt}%(amsalpha)
%\bibliography{../bib/quantum}

\end{document}